\documentclass[prb,twocolumn,superscriptaddress]{revtex4}

\usepackage{graphicx}
\usepackage{comment}
\usepackage{amsmath,amssymb,amsthm} 
\usepackage{verbatim}
\usepackage{float}
\usepackage{color}

\setcitestyle{super}

\usepackage[sort&compress]{natbib}

\begin{document}

\title{Impact of size polydispersity on the nature of \\ Lennard-Jones liquids}
\author{Trond S. Ingebrigtsen}
\email{trond@iis.u-tokyo.ac.jp}
\affiliation{Institute of Industrial Science, University of Tokyo, 4-6-1 Komaba, Meguro-ku, Tokyo 153-8505, Japan}
\author{Hajime Tanaka}
\email{tanaka@iis.u-tokyo.ac.jp}
\affiliation{Institute of Industrial Science, University of Tokyo, 4-6-1 Komaba, Meguro-ku, Tokyo 153-8505, Japan}
\date{\today}

\begin{abstract}
Polydisperse fluids are encountered everywhere in biological and
industrial processes. These fluids naturally show a rich phenomenology
exhibiting fractionation and shifts in critical point and freezing temperatures. 
Here, we study the impact of size polydispersity on the basic nature of
Lennard-Jones (LJ) liquids, which represent most molecular liquids 
without hydrogen bonds, via two- and three-dimensional molecular dynamics computer simulations. 
A single-component liquid constituting spherical particles and interacting
via the LJ potential is known to exhibit strong correlations between
virial and potential energy equilibrium fluctuations at constant
volume. This correlation significantly simplifies the physical
description of the liquid, and these liquids are 
now known as Roskilde-simple (RS) liquids. 
We show that this simple nature of the single-component LJ liquid is preserved even for 
very high polydispersities (above 40\% polydispersity for the studied uniform distribution). We also investigate isomorphs of moderately polydisperse LJ liquids. 
Isomorphs are curves in the phase diagram of RS liquids along which structure, dynamics, and some 
thermodynamic quantities are invariant in dimensionless units. We find
that isomorphs are a good approximation even for polydisperse LJ liquids. 
The theory of isomorphs thus extends readily to multi-component systems and can be used to 
improve even further the understanding of these intriguing systems.
\end{abstract}

\maketitle

\section{Introduction}

Polydisperse fluids are ubiquitous in both biological and industrial processes\cite{bagchi2012,wolynes2012}. 
Examples of polydisperse systems are: plastic materials, liquid foams, asphalt, micelles, and more. The property that typically
varies among the constituent molecules is the charge, mass, and size. The fluids' dispersity naturally induces a rich phenomenology not encountered 
in their single-component counterparts. As a consequence, polydisperse fluids have been the focus of a number of studies both 
theoretically \cite{dickinson1978,blum1979,salacuse1982,gualtieri1982,ginoza1997,evans1999,sollich2002,fasolo2003,jacobs2013} 
and via computer simulations \cite{frenkel1986,kofke1986,kofke1987,stapleton1990,auer2001,kristof2001,murarka2003,wilding2005,wilding2005b,wilding2006,kawasaki2007,abraham2008,wu2012,jacobs2013,sarkar2013,
ogarko2013,will2013,ashton2013,nguyen2014,phillips2014,sarkar2014}, as well as in 
experiments \cite{koningsveld1971,cowell1982,weeks2000,ye2005,watanabe2008,ballesta2008,banerjee2012,sacanna2013,palberg2014}.

These investigations revealed that polydisperse fluids exhibit a very rich phase diagram with shifts in critical point and freezing 
temperatures, as well as many-solid/fluid coexistence
regions \cite{stapleton1988,fasolo2003}. The coexisting phases in
fact show fractionation, i.e., the dispersed variable displays a distribution different from the parent in each phase. Moreover, polydisperse fluids exhibit  
terminal polydispersity beyond which the solid is not stable \cite{fasolo2003,sarkar2013,sarkar2014}. 
Polydispersity is thus regarded as an important factor to improve the
glass-forming ability of liquids and to control the 
liquid fragility \cite{tanakaR,kawasaki2007,watanabe2008,tanaka2010,abraham2008,abraham2008suppression}. 

Predicting the behavior of polydisperse fluids is a complex task, and several theoretical approaches have 
emerged to face the challenges of polydisperse fluids (see, e.g., Refs. \onlinecite{evans1999}, \onlinecite{sollich2002}, \onlinecite{jacobs2013}, and \onlinecite{ogarko2013}). 
The complex nature of the problem is enhanced by the fact that quasi-universal 
relations\cite{rosenfeld1,rosenfeld2,dzugutov1996} for monatomic fluids, possessing great predictive power, break down in increasingly 
polydisperse fluids \cite{pond2011,sarkar2014}. One such example is
Rosenfeld's excess entropy scaling\cite{rosenfeld1,rosenfeld2} in
which a dimensionless transport coefficient is correlated to the
excess entropy with respect to an ideal gas. The latter observation for polydisperse fluids stands in contrast to confined 
fluids where the quasi-universal relations are very robust as the degree of confinement is increased \cite{hsconfined,LJconfinedsmoothwalls,watanabe2011,ingebrigtsen2013, douglas2014}.

Explanations for quasi-universal relations in monatomic fluids were recently presented from a constant-potential-energy \textit{NVU}-dynamics 
perspective \cite{quasi,nvu1,nvu2,nvu3}, as well as from an exponential
pair potential perspective \cite{bacher2014,bacher2014n}. In these perspectives, the quasi-universal 
relations are attributed to the monatomic fluids being so-called Roskilde-simple \cite{paper1,paper2,paper3,paper4,paper5,thermoscl,prx}. 

Roskilde-simple (RS) liquids have been detailed in a number of papers \cite{paper1,paper2,paper3,paper4,paper5} and appear to be simpler than other types of liquids \cite{prx}.
RS liquids are characterized by having strong correlations between virial and potential energy equilibrium fluctuations at 
constant volume. In simulations of model liquids, the single-component Lennard-Jones (SCLJ) liquid, among many others, has been identified to belong to the 
class of RS liquids. Some experimental liquids have also been identified as RS \cite{gammagamma,roed2013,wence2014}.
More specifically, the class of RS liquids is believed to include most or all van der Waals and metallic liquids, but to exclude most 
or all covalent-bonding, hydrogen-bonding, strongly ionic, and dipolar liquids\cite{paper1}. Moreover, RS liquids are  
characterized by having isomorphs. Isomorphs are curves in the phase diagram of RS liquids along which structure, dynamics, and some 
thermodynamic quantities are invariant in dimensionless units; a
property that is not limited to the stable liquid region but also valid
in the supercooled liquid and even out-of-equilibrium\cite{paper4,sllod}.

Recently, Bagchi and coworkers\cite{sarkar2013,sarkar2014} studied Lindemann's melting and Hansen-Verlet's freezing criterion
in size polydisperse LJ liquids. The authors showed surprisingly that terminal polydispersity can be analyzed in terms of these rules. It
was also shown that order-parameter maps for these liquids are not independent but highly correlated. 
We note that melting and freezing lines of RS liquids are 
isomorphs from which the rules of Lindemann and Hansen-Verlet
follow\cite{paper4}. In fact, correlated order-parameter maps also follow from the existence
of isomorphs in the phase diagram\cite{paper4}.  

The observations of Bagchi and coworkers imply that polydisperse LJ liquids
retain the character of RS liquids. Conversely, the
breakdown of quasi-universal relations in polydisperse fluids may imply that polydisperse LJ liquids do
not remain RS. Thus, we investigate in the following the effect of size polydispersity on
strong virial-potential energy correlation in a systematic way to shed light on  
melting and freezing rules and quasi-universal behavior in polydisperse fluids.  
We note that previous investigations of multi-component RS liquids focused only on binary mixtures\cite{paper1,paper5,prx}.

Section \ref{sim} describes the applied methods and simulated model systems. Section \ref{ros} introduces RS liquids and explain their characteristics. 
In Sec. \ref{res} we show results from simulating size polydisperse LJ liquids in both two and three 
dimensions, in particular, the effect of size polydispersity on strong virial-potential energy correlation. 
Section \ref{dis} discusses the physics behind our observations. 
Finally, Sec. \ref{con} summarizes our work and presents an outlook.

\section{Simulation and model details}\label{sim}

We apply \textit{NVT} molecular dynamics computer simulations\cite{nose,hoover,nvttoxvaerd} to study size polydisperse LJ liquids in two and three dimensions. 
The RUMD package (see http://rumd.org) is used for three-dimensional simulations while an in-house code is used for two-dimensional simulations.
Pros and cons against applying constant-particle-number simulations to study polydisperse fluids are discussed in Ref. \onlinecite{wilding2010}.

The polydisperse LJ liquid provides a convenient framework for initiating theoretical investigations of polydispersity\cite{murarka2003,sarkar2014}, but has also been 
applied to mimic the behavior of for instance hydrocarbon or polymer solutions, idealized models for spherical micelles in water
above the critical micelle concentration, and polystyrene lattices stabilized with polyoxyethylene chains\cite{stapleton1988}. 

If the total potential energy of the system is given by $U$ = $\sum_{i<j}v(r_{ij})$, the size polydisperse LJ pair potential is
\setlength{\abovedisplayskip}{15pt}
\setlength{\belowdisplayskip}{10pt}
\begin{equation}\label{lj}
  v(r_{ij}) = 4\epsilon\Big[\Big(\frac{\sigma_{i j}}{r_{i j}}\Big)^{12} - \Big(\frac{\sigma_{i j}}{r_{i j}}\Big)^{6}\Big],
\end{equation}
where $\sigma_{i j}$ and $\epsilon$ set, respectively, the length and energy scale of the pair interaction between particle $i$ and particle $j$ ($i,j = 1, ..., N$, in 
which $N$ is the number of particles). As the study is focused on size polydisperse LJ liquids: $\epsilon$ is independent of the pair 
interaction and fixed in simulations. The parameter $\sigma_{i j}$ 
is defined from the Lorentz-Berthelot mixing rule\cite{tildesley} given by
\setlength{\abovedisplayskip}{15pt}
\setlength{\belowdisplayskip}{10pt}
\begin{align}
  \sigma_{i j} & = (\sigma_{i} + \sigma_{j})/2.
\end{align}
In this study, $\sigma_{i}$ and $\sigma_{j}$ are evenly distributed on an interval from $\sigma^{min}$ to $\sigma^{max}$. 
The polydispersity of the liquid $\delta$ is defined by 
\setlength{\abovedisplayskip}{15pt}
\setlength{\belowdisplayskip}{10pt}
\begin{align}
  \delta \equiv \sigma^{STD}(\sigma_{i})/\bar{\sigma}_{i} = \frac{1}{\sqrt{3}} \, \frac{\sigma^{max}-\sigma^{min}}{\sigma^{max}+\sigma^{min}},
\end{align}
where $\sigma^{STD}(\sigma_{i})$ and $\bar{\sigma}_{i}$ are,
respectively, the dispersed variable's standard deviation and mean.
We study polydispersities in the range $\delta$ = 0\% to $\delta$ =
52\% and apply a truncated-and-shifted pair potential cutoff at $r_{c} = 2.5\sigma_{i j}$.
We note that care should be taken when comparing the absolute polydispersity number $\delta$ amongst different distributions\cite{murarka2003}. 

\begin{figure}[H]
  \centering
  \includegraphics[width=70mm]{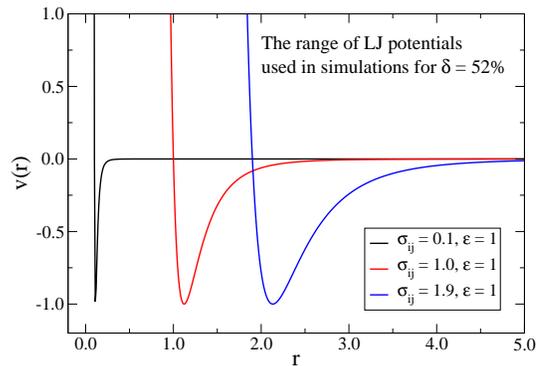}
  \caption{The range of LJ potentials used in simulations of a polydisperse LJ liquid with $\delta$ = 52\%. }
  \label{pair}
\end{figure}

The units of the simulation are defined by setting the mean values
over the $N$ particles to unity, i.e., $\sigma_{\bar{N} \bar{N}} = 1$, $\epsilon_{\bar{N} \bar{N}} = 1$, $m_{\bar{N}}$ = 1. 
All particle masses are identical (and equal to unity). 
Figure \ref{pair} shows pair potentials of Eq. (\ref{lj}) with
$\sigma_{ij}$ = 0.1, 1.0, 1.9 and $\epsilon$ = 1. The latter
corresponds to the range of LJ potentials used
in simulations of a size polydisperse LJ liquid with $\delta$ = 52\% (i.e., $\sigma^{min}$ = 0.1 and $\sigma^{max}$ = 1.9). For this high polydispersity the simulated pair potentials cover a very broad range of potentials.
\newline \newline 

The two-dimensional simulations use $N$ = 1000 particles, whereas the three-dimensional simulations use $N = 5000$ particles. The finite-size effects 
are larger in two dimensions than in three dimensions. In either case,
however, finite-size effects are observed to be small for the
quantities studied in this paper and no additional averaging 
is performed.

\section{Roskilde-simple liquids}\label{ros}

RS liquids and their associated simple properties have been detailed in a number of papers, starting with a series of five papers\cite{paper1,paper2,paper3,paper4,paper5}.
RS liquids are characterized by having strong correlations between virial $W$ and potential energy $U$ equilibrium fluctuations at 
constant volume. The virial-potential energy correlation is quantified via Pearson's correlation coefficient $R$ given by 
\setlength{\abovedisplayskip}{15pt}
\setlength{\belowdisplayskip}{10pt}
\begin{equation}\label{R}
  R = \frac{\langle \Delta W \Delta U \rangle}{\sqrt{\langle (\Delta W)^{2} \rangle}\sqrt{ \langle (\Delta U)^{2} \rangle}},
\end{equation}
in which $\Delta$ is deviation from mean values, $\langle ... \rangle$ denotes \textit{NVT} ensemble averages, 
and $-1$ $\leq$ $R$ $\leq$ $1$. RS liquids are defined pragmatically by requiring $R$ $\geq$
$0.90$. $R$ depends on the state point investigated but has been shown to be high in large parts of 
the phase diagram or not at all. 

Only inverse power-law (IPL) fluids with $r^{-n}$ pair potentials are perfectly correlating ($R$ = 1
since $\Delta W = (n/3)\Delta U$), but many 
models\cite{paper1,prx,veldhorst2014}, such as the SCLJ liquid, the Kob-Andersen binary LJ mixture,  
the Lewis-Wahnstr{\"o}m OTP model, the asymmetric dumbbell model,
identical chain length LJ-bead polymer models, and some experimental liquids\cite{gammagamma,roed2013,wence2014}, have been 
shown to belong to the class of RS liquids. 

The origin behind strong correlation in the SCLJ liquid is as follows.
The LJ pair potential can be approximated by an
  IPL potential in the repulsive
  part\cite{coslovich1,coslovich2,coslovich3}, i.e., below the
  potential mininum with an IPL-exponent $n \approx 18$. The required
  exponent is larger than the repulsive term $r^{-12}$ encountered in
  the LJ potential
  as the effect of attraction is to increase the steepness of the
  repulsive part. However, contributions to the fluctuations
  come from pair distances around
  the pair potential minimum\cite{paper2,prx}. As a consequence,
  one must add a linear term and a constant to the IPL potential to obtain a proper
  description of the LJ potential: 
  \begin{equation}
    v_{LJ}(r) \approx Ar^{-n} + B + Cr. \label{eq:potential}
  \end{equation}
The constraint of constant volume in the \textit{NVT} ensemble has the following 
effect: When one nearest-neighbor distance increases another one 
decreases. Thus, upon summation the contribution from the linear term
to $U$ and $W$ is almost constant and vanishes with respect to
the fluctuations. For a very recent review of RS liquids, see Ref. \onlinecite{dyre2014}.

As mentioned, RS liquids are associated with a number of simple properties. RS liquids are, for instance,
characterized by having isomorphs to a good approximation\cite{paper4}. Consider two state points in a liquid's  
phase diagram with density and temperature ($\rho_{1}$, $T_{1}$) and ($\rho_{2}$, $T_{2}$). These two state points are defined to 
be isomorphic if the following holds: Whenever a microconfiguration of 
state point ($1$) and of state point ($2$) have the same 
reduced coordinates, $\rho_{1}^{1/3}$ $\textbf{R}^{(1)}$ = $\rho_{2}^{1/3}$ $\textbf{R}^{(2)}$ ($\textbf{R}$ denotes a 3$N$-dimensional configurational-space vector and $\rho \equiv N/V$, in which $V$ is the 
volume),
these two configurations have proportional Boltzmann factors, i.e.,
\begin{equation} \label{defiso}
  e^{-U(\textbf{R}^{(1)})/k_{B}T_{1}} = C_{12} \, e^{-U(\textbf{R}^{(2)})/k_{B}T_{2}}.
\end{equation}
Here $C_{12}$ is a constant and depends only on the state points ($1$) and ($2$). An isomorph is  
defined as a continuous curve of state points that are all pairwise
isomorphic. For IPL fluids $C_{12}$ = 1, and isomorphs
  thus encapsulate the well-known scaling properties\cite{hoover1970} of these
  systems. However, for all other RS liquids $C_{12} \neq 1$; accordingly, the state points can be represented  
 effectively on a one-dimensional phase diagram only for certain properties \cite{paper4}.

An immediate consequence of isomorphs is that a number of quantities are invariant 
along these curves. The invariants include the structure and dynamics
in reduced units and some thermodynamic quantities, e.g., the excess
entropy with respect to an ideal gas $s_{ex}\equiv S_{ex}/N$.
We note that reduced units refer to macroscopic quantities, such as $\rho^{-1/3}$ and $T$, rather than the usual approach in simulations via 
microscopic quantities (see Ref. \onlinecite{paper4} for details).

For RS liquids, temperature separates\cite{thermoscl} in a function of density and of excess entropy: $k_{B}T = h(\rho)f(s_{ex})$. 
From this equation, one observes the isomorph invariant $h(\rho)/T$ which was used to define the so-called isomorph scaling\cite{thermoscl,beyond}, 
in which the reduced relaxation time $\tilde{\tau}_{\alpha}$ is a function $\tilde{\tau}_{\alpha}$ = $f(h(\rho)/T)$. Isomorph 
scaling explains classical density scaling\cite{reviewRoland} for which $\tilde{\tau}_{\alpha} = f(\rho^{\gamma}/T)$, where $\gamma$ is a fitting exponent. 
Density scaling is valid only for relatively small density variations, whereas isomorph scaling works even  
when the density variation is large\cite{beyond}. Additionally, the isomorph invariant $h(\rho)/T$ provides a convenient way to generate isomorphs since $h(\rho)$ inherits the analytical structure of the potential energy function\cite{thermoscl}.

There exist additional simple observations for RS liquids than mentioned so far, e.g., the 
dominance of first-coordination shell interactions\cite{prx,ingebrigtsen2014}, extension to non-trivial boundary conditions such as nanoconfinement\cite{ingebrigtsen2013} or shear-flow\cite{sllod}, 
crystalline isomorphs\cite{crystals}, and the relevance of Rosenfeld-Tarazona's expressions\cite{RT} for these liquids\cite{RTIngebrigtsen}. 
The interested reader is referred to these and other papers for more information.

Finally, a reformulation of isomorphs was very recently presented\cite{thomas2014}. If $\textbf{R}_{a}$ and $\textbf{R}_{b}$ denote two different 3$N$-dimensional 
configurational-space vectors of a chosen state point, the isomorphic condition is
\begin{equation}
  U(\textbf{R}_{a}) < U(\textbf{R}_{b}) \Rightarrow U(\lambda\textbf{R}_{a}) < U(\lambda\textbf{R}_{b}),
\end{equation}
where $\lambda$ is a scalar. This definition encapsulates the previous definition of isomorphs as a (very good) first-order 
approximation\cite{thomas2014}. The mentioned invariants are exact
within this reformulation, too. We note, however, that the constant 
volume heat capacity is not formally invariant.

\section{Results from polydisperse LJ liquids}\label{res}

The simulations in this paper were performed both in two and three dimensions, and
thus we discuss each case separately beginning with the two-dimensional results. 

\subsection{Simulations in two dimensions}

The investigation is initiated by considering the effect of size polydispersity on the structure and dynamics 
of the SCLJ liquid in two dimensions at $\rho$ = 0.75 and $T$ = 0.70. For 0\% polydispersity, this state point corresponds to the liquid phase.

\begin{figure}[H]
  \centering
  \includegraphics[width=70mm]{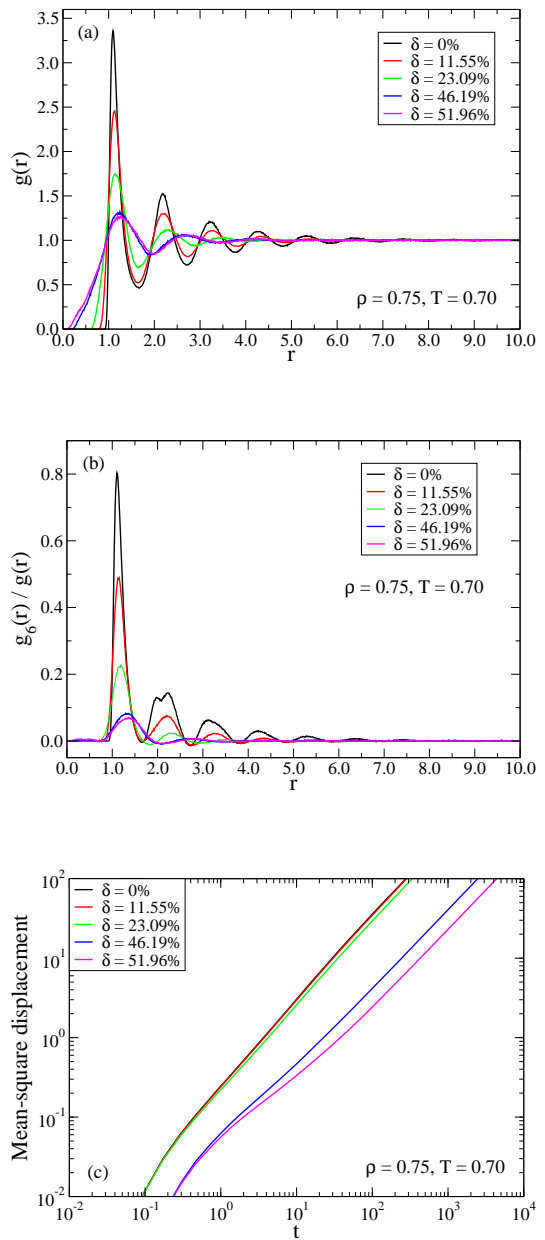}
  \caption{Structural and dynamical effects of size polydispersity on the SCLJ liquid in two dimensions at $\rho$ = 0.75 and $T$ = 0.70. (a) Radial distribution functions (RDFs). 
    (b) Bond-orientational-order functions (BOFs, see
    Eq. (\ref{bond2D})). (c) Mean-square displacements (MSDs). Size polydispersity is seen to diminish 
    structural order and slow down dynamics at constant density and
    temperature.}
  \label{polyeffect}
\end{figure}

Figure \ref{polyeffect}(a) presents radial distribution functions
(RDFs) upon the increase of polydispersity from $\delta$ = 0\% to $\delta$ = 52\%,
whereas Fig. \ref{polyeffect}(b) shows bond-orientational-order functions\cite{binder2002,kawasaki2007} (BOFs) defined via
\setlength{\abovedisplayskip}{15pt}
\setlength{\belowdisplayskip}{10pt}
\begin{equation}\label{bond2D}
  g_{6}(r) \equiv \frac{L^{2}}{2 \pi r \Delta r N(N-1)}\sum_{i \neq k}\delta\big(r-|\textbf{r}_{ik}|\big)\psi^{i}_{6}\,\psi^{k*}_{6}.
\end{equation} 
The division with $g(r)$ in Fig. \ref{polyeffect}(b) approximately accounts for effects due to positional order in the liquid\cite{steinhardt1983,kawasaki2007}.
In the equation above, $\delta$ denotes Dirac's $\delta$-function, $\textbf{r}_{ik}$ $\equiv$ $\textbf{r}_{i} - \textbf{r}_{k}$, $L$ is the 
box length, and $\psi^{i}_{6}$ $\equiv$ $1/n_{i}\sum_{m=1}^{n_{i}}e^{j6\theta_{m}^{i}}$, in which $n_{i}$ is the
number of nearest neighbors of particle $i$, and $\theta_{m}^{i}$ is the
angle between the vector $\textbf{r}_{mi}$ and the $x$-axis (particle $m$ is a neighbor of particle $i$). Nearest-neighbor particles are identified via a 
Voronoi-construction\cite{tildesley}. Finally, Fig. \ref{polyeffect}(c) shows mean-square displacements (MSDs). 

Both measures of structural order (Figs. \ref{polyeffect}(a) and (b)) are seen to vanish as polydispersity is increased from $\delta$ = 0\% to $\delta$ = 52\% at constant density and temperature.
The MSD shows approximately one order-of-magnitude decrease in the
diffusion coefficient over the same variation in
polydispersity. Introducing size polydispersity has thus a pronounced effect on the SCLJ liquid. The latter results confirm several previous simulations of polydisperse hard-sphere and LJ
liquids\cite{frenkel1986,murarka2003,sarkar2013}. We note that if the volume (area) fraction $\phi$ $\equiv$ $1/L^{2}\sum_{i}\pi (\sigma_{i}/2)^{2}$ of the system is 
kept constant, instead of density, a similar diminishment of structural
correlations is observed. However, in the case of the dynamics an enhancement of the mobility is seen\cite{abraham2008}.
\newline \newline 
\begin{figure}[H]
  \centering
  \includegraphics[width=70mm]{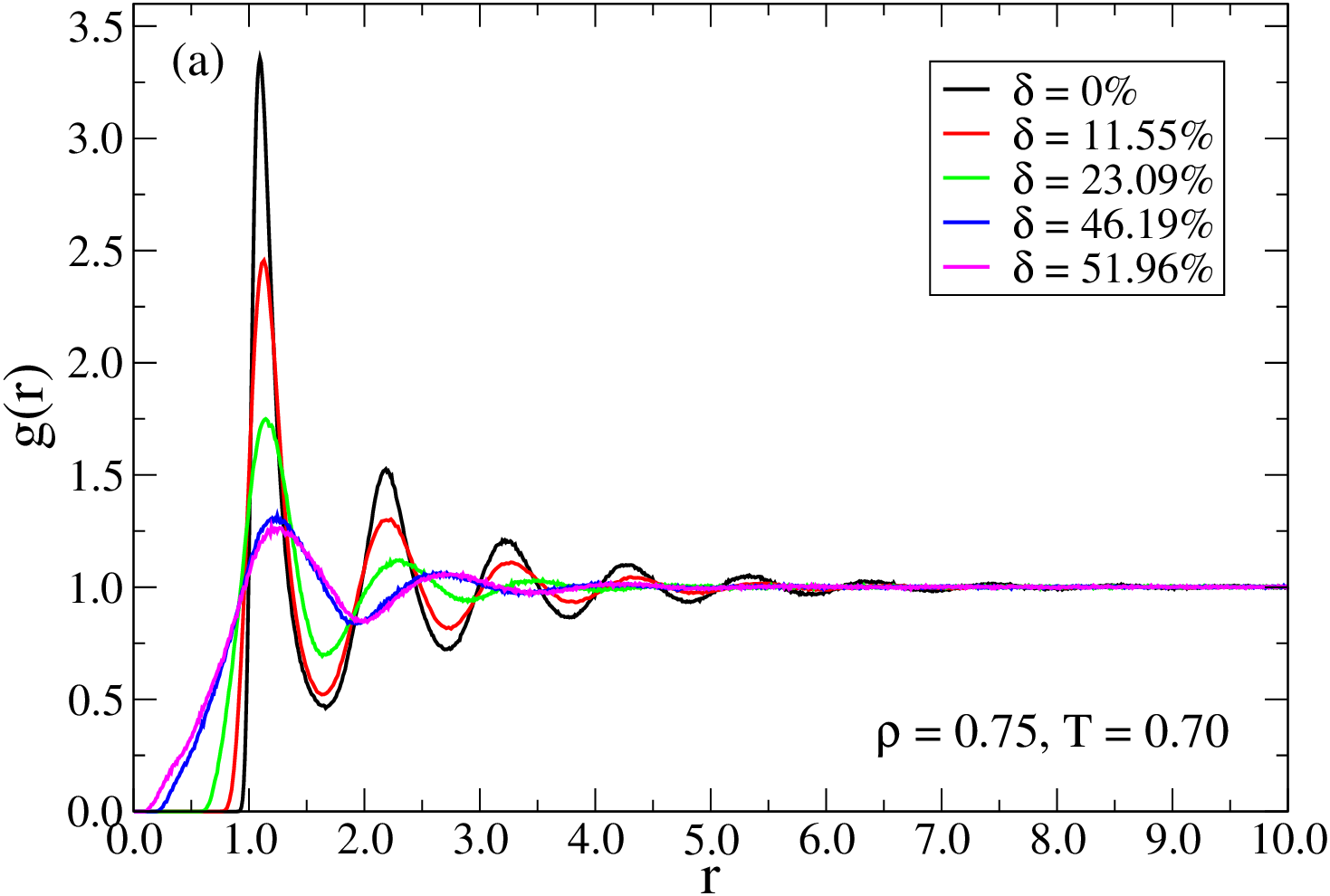}
\end{figure}
\begin{figure}[H]
  \centering
  \includegraphics[width=70mm]{rdf6}
\end{figure}

\begin{figure}[H]
  \centering
  \includegraphics[width=68mm]{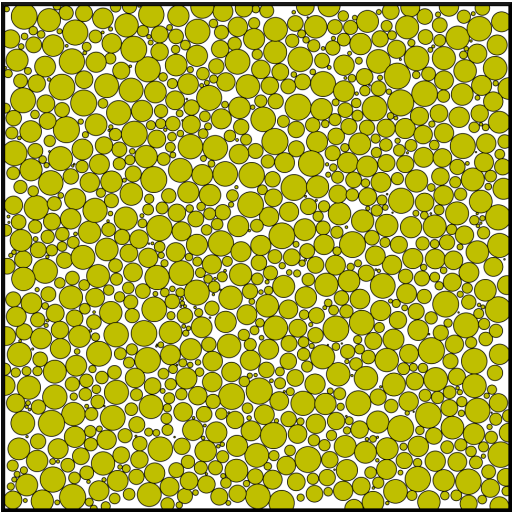}
  \caption{A snapshot of a particle configuration ($N$ = 1000) for a highly 
    size polydisperse LJ liquid with $\rho$ = 0.75, $T$ = 0.70, $\delta$ = 52\%. Each particle is represented according to 
    its value of $\sigma_{i}$: As the LJ potential is
    continuous some of the disks overlap slightly. A multitude of different sizes are
    present at this high polydispersity.}
  \label{2dplot}
\end{figure}

Furthermore, Fig. \ref{2dplot} shows a snapshot of a particle configuration for the size polydisperse 
LJ liquid with $\rho$ = 0.75, $T$ = 0.70, $\delta$ = 52\%. Each particle is represented according to its value of $\sigma_{i}$. A multitude of 
different sizes are present at this high polydispersity exemplifying again
the rich and non-trivial diversity of the liquids studied. 
\newline \newline 

The study is now turned towards the assessment of the effect of size polydispersity from the viewpoint of RS liquids. 
In this connection, we note that previous simulations of RS liquids were performed only in three dimensions. 
The simulations in this section are two-dimensional.

\begin{figure}[h]
  \centering
  \includegraphics[width=80mm]{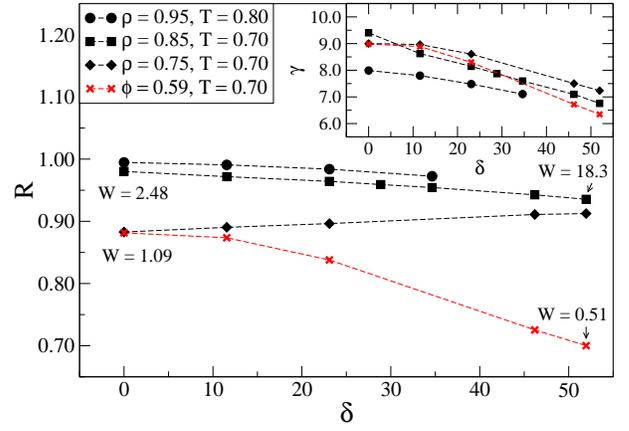}
  \caption{Effect of polydispersity $\delta$ on the correlation coefficient $R$ (main figure, see Eq. (\ref{R})) and density-scaling 
    exponent $\gamma$ $\equiv$ $\langle \Delta U \Delta W \rangle/\langle (\Delta U)^{2}\rangle$ (inset) for the polydisperse LJ
    liquid in two dimensions. Black data points are for constant density $\rho$, and red data points are for constant volume (area) fraction $\phi$.
    For 0\% polydispersity, $\rho$ = 0.85 is inside the coexistence region, and $\rho = 0.95$ is inside the crystalline region.}
  \label{poly2d}
\end{figure}

Figure \ref{poly2d} shows the correlation coefficient $R$ as function of polydispersity $\delta$ at both
constant density $\rho$ (black data points) and constant
volume fraction $\phi$ (red data points). For the three-dimensional SCLJ liquid, a typical value for the correlation coefficient is $R \approx 0.95$.
The inset shows the density-scaling exponent\cite{paper4,thermoscl} $\gamma$ $\equiv$ $\langle \Delta U \Delta W \rangle/\langle (\Delta U)^{2}\rangle$ as function of
polydispersity (see Sec. \ref{ros}).
\newline \newline 

At constant density, we observe both a decrease and an increase in $R$ depending on the starting state point at 0\% polydispersity. 
This observation is expected as the phase diagram for polydisperse liquids shows very rich phenomenology\cite{stapleton1988,fasolo2003}. 
Although a decrease in $R$ is observed for the two highest densities ($\rho$ = 0.95 and $\rho$ = 0.85) all state points investigated are RS.
For the lowest density ($\rho$ = 0.75) the liquid becomes RS at high polydispersities. In fact, the results show that
polydisperse LJ liquids at 52\% polydispersity are RS liquids. This observation is non-trivial 
given the complex nature of polydisperse liquids (see also Fig. \ref{2dplot}).

In the case of constant volume fraction, however, we observe a significant decrease in $R$ as function of polydispersity. An increase in polydispersity, at constant volume
fraction, is accompanied  by a decrease in pressure. This is in
contrast to constant density which increases the pressure; Table
\ref{table} shows selected polydispersities and corresponding volume fractions along the
three simulated constant-density curves. For the density-scaling exponent $\gamma$, we only observe a decrease in its value,
when polydispersity is increased.

\begin{figure}[h]
  \centering
  \includegraphics[width=70mm]{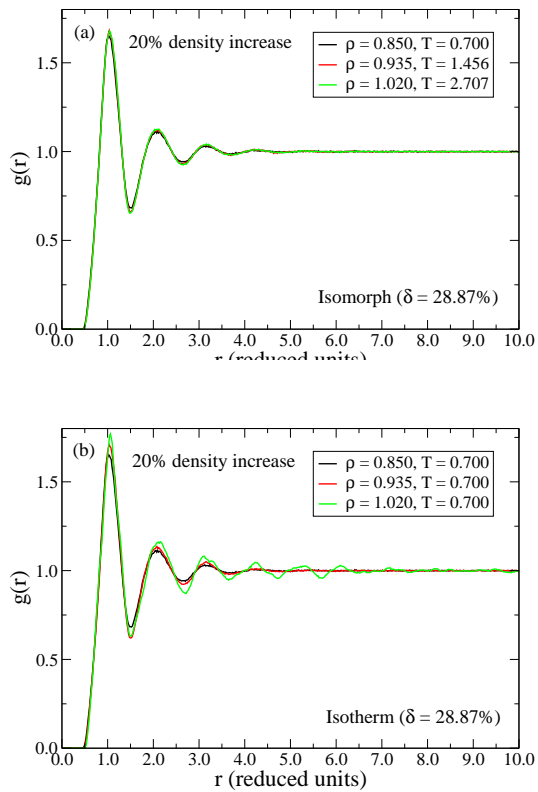}
  \caption{RDFs for a moderately polydisperse LJ liquid ($\delta$ =
    28.87\%) in two dimensions. (a) Along an isomorph. (b) Along an
    isotherm. The highest density state point is not fully equilibrated.}
  \label{rdf2d}
\end{figure}

\begin{table}[h]
\begin{tabular}
{|p{1.0cm}||p{2.0cm}|p{2.0cm}|p{2.0cm}|}
\hline
\hline
$\delta$ & $\phi$ ($\rho$ = 0.75) & $\phi$ ($\rho$ = 0.85) & $\phi$ ($\rho$ = 0.95) \\
\hline
\hline
0     & 0.589 & 0.668 & 0.746 \\
11.55 & 0.597 & 0.676 & 0.755 \\
23.09 & 0.619 & 0.702 & 0.784 \\
34.64 & -     & 0.746 & 0.834 \\
51.96 & 0.746 & 0.845 & - \\
\hline
\hline
\end{tabular}
\caption{Selected polydispersities $\delta$ and corresponding volume (area) fractions $\phi$ along the
three simulated constant-density curves.}
\label{table}
\end{table}
As mentioned previously, the quantity $h(\rho)/T$ is invariant along an isomorph (see Sec. \ref{ros}). The function $h(\rho)$ for a LJ liquid in two dimensions is given by 
\setlength{\abovedisplayskip}{15pt}
\setlength{\belowdisplayskip}{10pt}
\begin{equation}\label{h}
  h(\tilde{\rho}) = (\gamma_{*}/3 - 1)\tilde{\rho}^{6} + (2 - \gamma_{*}/3)\tilde{\rho}^{3}.
\end{equation}
In the equation above: $\tilde{\rho}$ $\equiv$ $\rho / \rho_{*}$,
$\rho_{*}$ is a chosen reference density, and $\gamma_{*}$ is the 
value of $\gamma$ = $\langle \Delta U \Delta W \rangle/\langle (\Delta U)^{2}\rangle$ obtained from the equilibrium fluctuations at $\rho_{*}$ 
(see Refs. \onlinecite{thermoscl} and \onlinecite{beyond} for
additional details). In this and the following section the reference
state point is $\rho_{*}$ = 1 and $T_{*}$ = 1.
We now proceed to study isomorphs of a moderately polydisperse LJ liquid. Isomorphs are generated 
by the procedure stated below\cite{thermoscl,beyond}:

\begin{enumerate}
\item A starting state point is chosen.
\item Density is varied by some percentage (typically 10-20\% but can be even a factor of two\cite{beyond}).
\item The temperature of the isomorphic state point is calculated via Eq. (\ref{h}) keeping $h(\rho)/T$ constant.
\item A simulation is performed at the predicted state point and the
  procedure is repeated.
\end{enumerate}
Figure \ref{rdf2d}(a) shows RDFs along an isomorph with 20\% density
increase ($\gamma_{*}$ = 7.02 and starting state point $\rho$ = 0.85 and $T$ = 0.70). For 
reference, Fig. \ref{rdf2d}(b) shows an isotherm with the same density increase. A good invariance of structure is seen along the isomorph, whereas 
structure on the isotherm clearly shows a poorer scaling.
\newline \newline 
\begin{figure}[h]
  \centering
  \includegraphics[width=70mm]{isomorph_rdf}
\end{figure}

\begin{figure}[h]
  \centering
  \includegraphics[width=70mm]{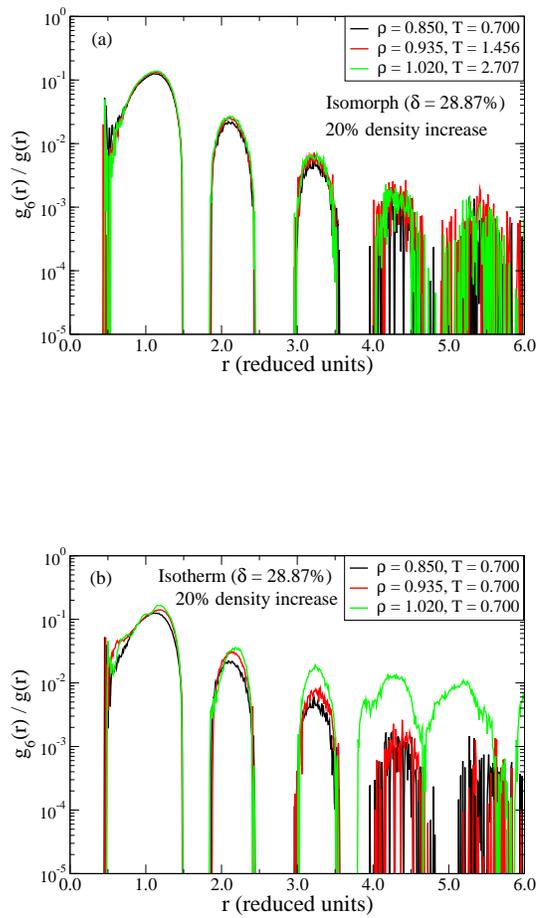}
  \caption{BOFs for a moderately polydisperse LJ liquid ($\delta$ = 28.87\%) in two dimensions. (a) Along an isomorph. (b) Along an isotherm. 
    The highest density state point is not fully equilibrated.}
  \label{g62d}
\end{figure}

Figure \ref{g62d} shows the corresponding BOF figures. A good invariance is again seen along the isomorph but not 
along the isotherm.
\newline \newline 
\begin{figure}[h]
  \centering
  \includegraphics[width=70mm]{isomorph_rdf6}
\end{figure}

\begin{figure}[h]
  \centering
  \includegraphics[width=70mm]{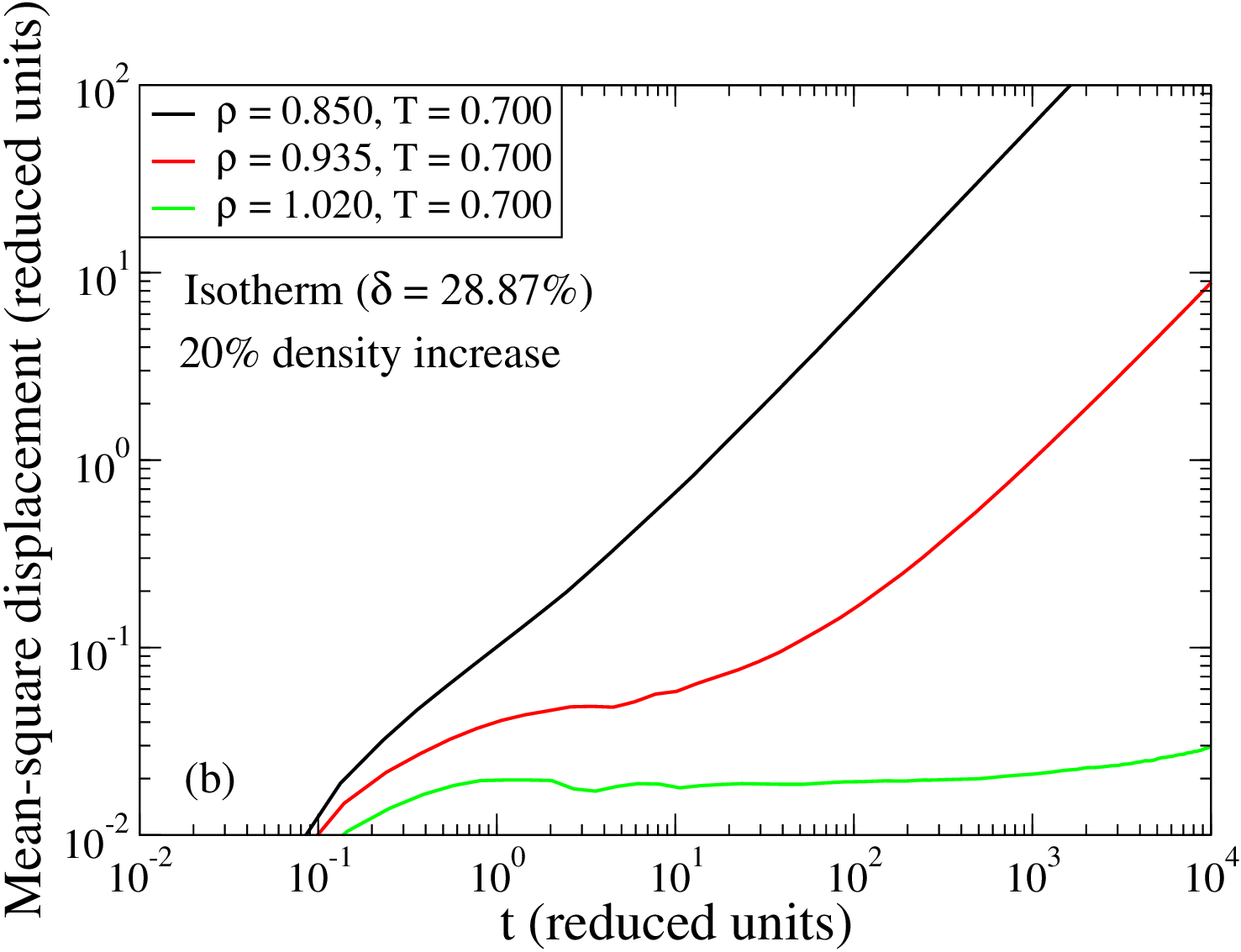}
  \caption{MSDs for a moderately polydisperse LJ liquid ($\delta$ =
    28.87\%) in two dimensions. (a) Along an isomorph. (b) Along an
    isotherm. The highest density state point 
    is not fully equilibrated.}
  \label{msd2d}
\end{figure}

Turning to the dynamics in Fig. \ref{msd2d}, a near-perfect collapse is seen for the MSD on the isomorph whereas, as expected, a poor invariance is seen along the isotherm. 
The fact that liquids are highly size polydisperse is thus not a hindrance to showing simple isomorphic behavior.
\newline \newline 
\begin{figure}[h]
  \centering
  \includegraphics[width=70mm]{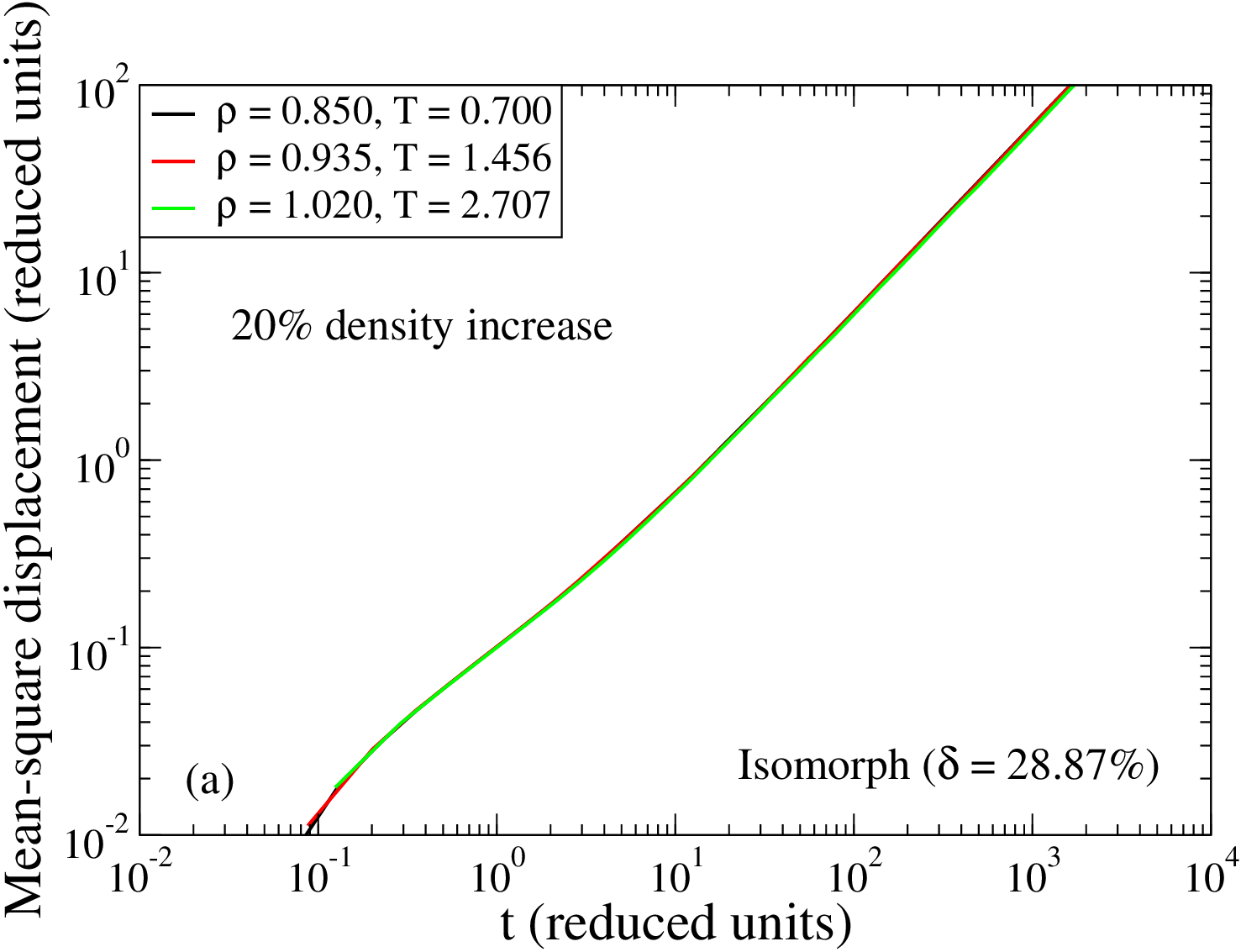}
\end{figure}

\subsection{Simulations in three dimensions}\label{three}

\begin{figure}[H]
  \centering
  \includegraphics[width=80mm]{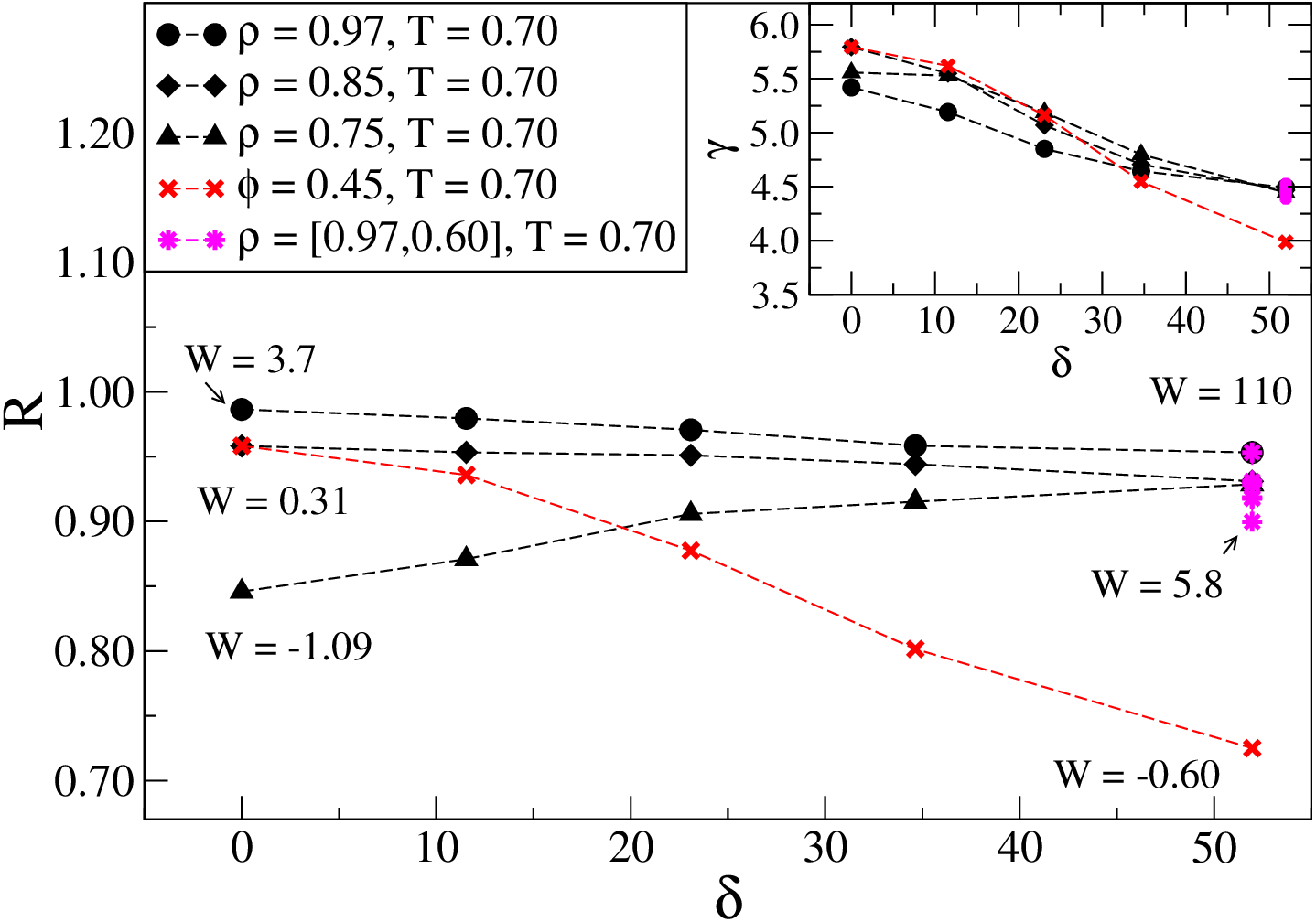}
  \caption{Effect of polydispersity $\delta$ on the correlation coefficient $R$ (main figure, see Eq. (\ref{R})) and density-scaling 
    exponent $\gamma = \langle \Delta U \Delta W \rangle/\langle (\Delta U)^{2}\rangle$ (inset) for the polydisperse LJ
    liquid in three dimensions. Black data points are for constant density $\rho$, and red data points are for
    constant volume fraction $\phi$. 
    The highest polydispersity data points of $\rho$ = 0.75 and $\rho$
    = 0.85, and the two highest polydispersity data points of $\rho$ = 0.97 are not fully 
    equilibrated due to very slow dynamics. For 0\% polydispersity:
    $\rho$ = 0.97 is inside the crystalline region. The magenta asteriks data
    points span densities ($T$ = 0.70): 0.97, 0.85, 0.75, 0.70, 0.65, and 0.60. }
  \label{R3d}
\end{figure}

This section studies size polydisperse LJ liquids in three dimensions. Figure \ref{R3d} shows the correlation coefficient
$R$ as function of polydispersity at constant density and volume fraction. As for the two-dimensional 
results both a decrease and an increase in $R$ is observed, when
polydispersity is increased at constant density (i.e., increasing pressure). For the density $\rho$ = 0.75, the liquid becomes RS at high polydispersities, whereas for
$\rho$ = 0.85 and $\rho$ = 0.97 all state points investigated are RS, albeit $R$ decreases. We observe again that highly size polydisperse LJ liquids are RS: a non-trivial observation.
\newline \newline 

In the case of constant volume fraction, a significant
decrease in $R$ is observed and is accompanied by the virial turning negative (but not the pressure). 
Actually, the liquid goes from RS to non-RS in the studied range of polydispersities. For single-component 
liquids: $R$ usually decreases rapidly, when the virial becomes negative\cite{paper1}. 

The density-scaling exponent $\gamma$ shows $-$ as in two dimensions $-$
a decrease in its value as function of polydispersity. Actually, it is observed from Fig. \ref{R3d} 
that all constant-density curves collapse at high polydispersities 
with $\gamma$ $\approx$ 4.5. Even a path at $\delta$ = 52\%, where
density varies from $\rho$ = 0.97 to 0.60 ($\phi$ = 0.94 to 0.58) and the virial from $W$ = 110 to 5.8, gives approximately 
the same value of $\gamma$ (see magenta asteriks data points). 

This observation indicates that for highly size polydisperse LJ liquids $\gamma$ is mainly
controlled by the degree of polydispersity $\delta$. 
In the non-reformulated isomorph theory $\gamma$ is predicted to be a function only of density\cite{thermoscl} (see Sec. \ref{ros}). For the SCLJ liquid, a value of
 $\gamma$ $\approx$ 5.5 is obtained at low-to-moderate pressures (see inset and Ref. \onlinecite{thermoscl}).

Very recently, an expression was established relating $\gamma$ to the
pair potential of monatomic RS liquids\cite{bohlingN2}. The expression is given by 
\begin{align}
  \gamma(\rho) & = \frac{n^{(2)}(r)}{3}\Big |_{r=\Lambda\rho^{-1/3}},\label{n2}\\
  n^{(2)}(r)  & \equiv -2 - r\frac{v^{(3)}(r)}{v^{(2)}(r)} \label{n2a},
\end{align}
where $\Lambda$ is a dimensionless number, and $v^{(p)}$ is the $p$'th derivative
of the pair potential $v$. $\gamma$ is then related to the curvature and torsion of the pair
potential. $\Lambda$ is chosen as the
reduced distance ($\tilde{r}$ = $r\rho^{1/3})$ at which $r^{2}g(r)$ obtains its maximum, i.e., the most likely nearest-neighbor distance\cite{bohlingN2}. In this case, however, $\gamma$ is not a function 
only of density. 

\begin{figure}[H]
  \centering
  \includegraphics[width=70mm]{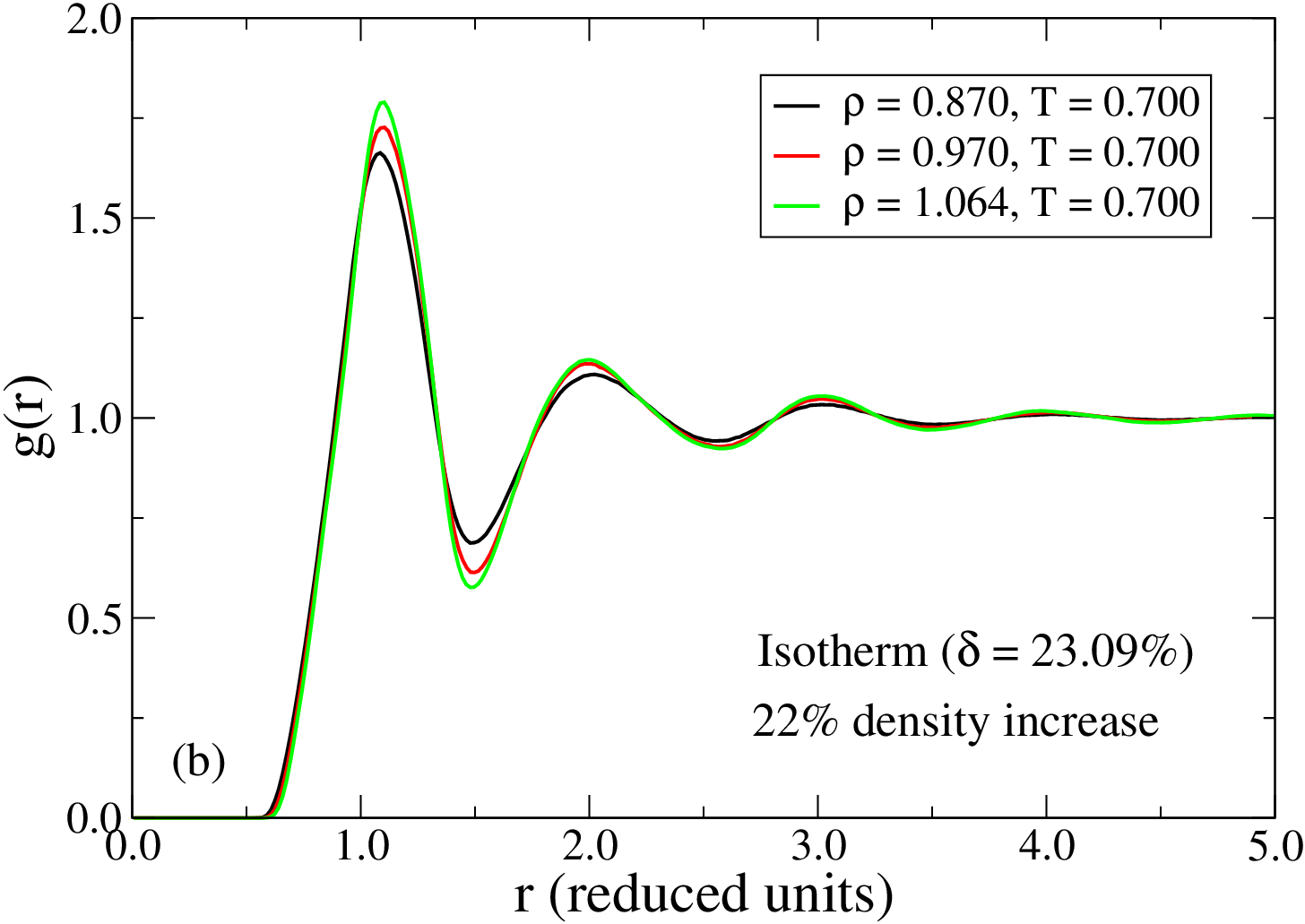}
  \caption{RDFs for a moderately polydisperse LJ liquid ($\delta$ =
    23.09\%) in three dimensions. (a) Along an isomorph. (b) Along an
    isotherm. The highest density state point 
    is not fully equilibrated.}
  \label{rdf3d}
\end{figure}

RDFs along an isomorph and isotherm for a polydisperse LJ liquid with
$\delta$ = 23.09\% are given in Fig. \ref{rdf3d} ($\gamma_{*}$ = 4.79 and 
starting state point $\rho$ = 0.97 and $T$ = 0.70). The isomorph is generated via the aforementioned 
procedure keeping $h(\rho)/T$ constant. The expression\cite{thermoscl} for $h(\rho)$ in three dimensions is given by
\setlength{\abovedisplayskip}{15pt}
\setlength{\belowdisplayskip}{10pt}
\begin{align}
  h(\tilde{\rho}) = (\gamma_{*}/2 - 1)\tilde{\rho}^{4} + (2 - \gamma_{*}/2)\tilde{\rho}^{2},
\end{align}
with definitions provided in the previous section. The RDF is to a good approximation invariant along the isomorph and less so on the 
isotherm (the highest density state point is, however, not fully equilibrated due to slow dynamics).
\newline \newline 
\begin{figure}[H]
    \centering
  \includegraphics[width=70mm]{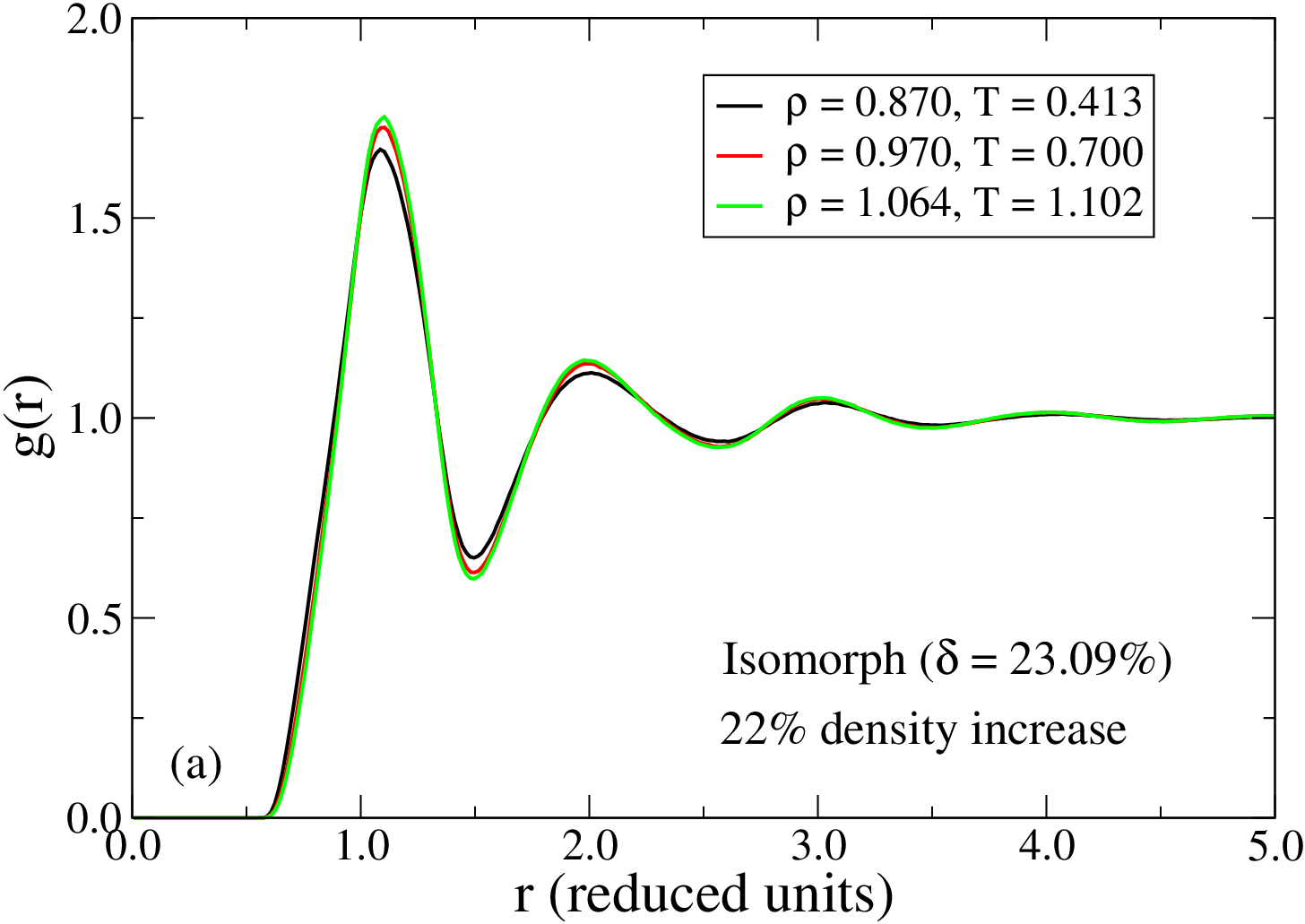}
\end{figure}

\begin{figure}[H]
  \centering
  \includegraphics[width=70mm]{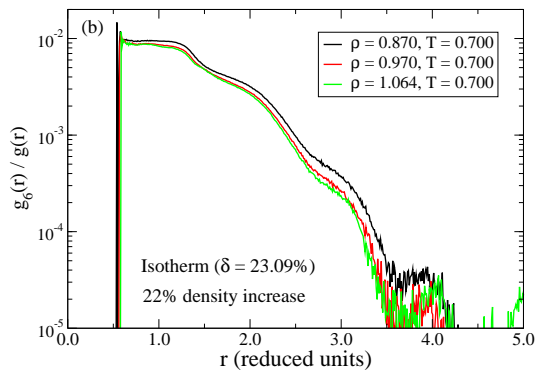}
  \caption{BOFs for a moderately polydisperse LJ liquid ($\delta$ =
    23.09\%) in three dimensions. (a) Along an isomorph. (b) Along an
    isotherm. The highest density state point 
    is not fully equilibrated.}
  \label{BOF3d}
\end{figure}

Figure \ref{BOF3d} shows BOFs along the same isomorph and isotherm. The BOF is defined in three dimensions via

\begin{equation}\label{bond}
  g_{6}(r) \equiv \frac{L^{3}}{4 \pi r^{2} \Delta r N(N-1)}\sum_{i \neq k}\delta\big(r-|\textbf{r}_{ik}|\big)S^{ik}_{6},
\end{equation}
in which 

\begin{align}
  S^{ik}_{6} & \equiv \frac{4\pi}{12 + 1}\sum_{m=-6}^{6}Q_{6m}^{i}Q_{6m}^{k*},\\
  Q_{6m}^{i} & \equiv \frac{1}{n_{i}+1}\sum_{g=1}^{n_{i}+1}\frac{1}{n_{g}}\sum_{j=1}^{n_{g}} Y_{6m}(\textbf{r}_{jg}). 
\end{align}
The above expression for $Q_{6m}^{i}$ gives coarse-grained Steinhardt bond-orientational-order parameters\cite{steinhardt1983,lechner2008}.
$Y_{6m}$ is the spherical harmonic function with degree $l$ = 6 and order $m$ = $[-6,6]$. The notation $n_{i}+1$ indicates that the sum is 
taken over all the nearest neighbors of particle $i$ including particle $i$ itself; nearest-neighbor particles are identified via 
the twelve-closest particles. From Fig. \ref{BOF3d}, a similar conclusion as for the RDF is reached; a good invariance is observed along the isomorph and less so on the isotherm.
These results also indicate a good invariance of the first coordination shell (FCS) environment along the isomorph and are consistent with the role 
of the FCS in determining the structure and dynamics of RS liquids (see Ref. \onlinecite{prx} for more information on FCS results).
\newline \newline 
\begin{figure}[H]
  \centering
  \includegraphics[width=70mm]{3D_rdf6_isomorph}
\end{figure}

\begin{figure}[H]
  \centering
  \includegraphics[width=70mm]{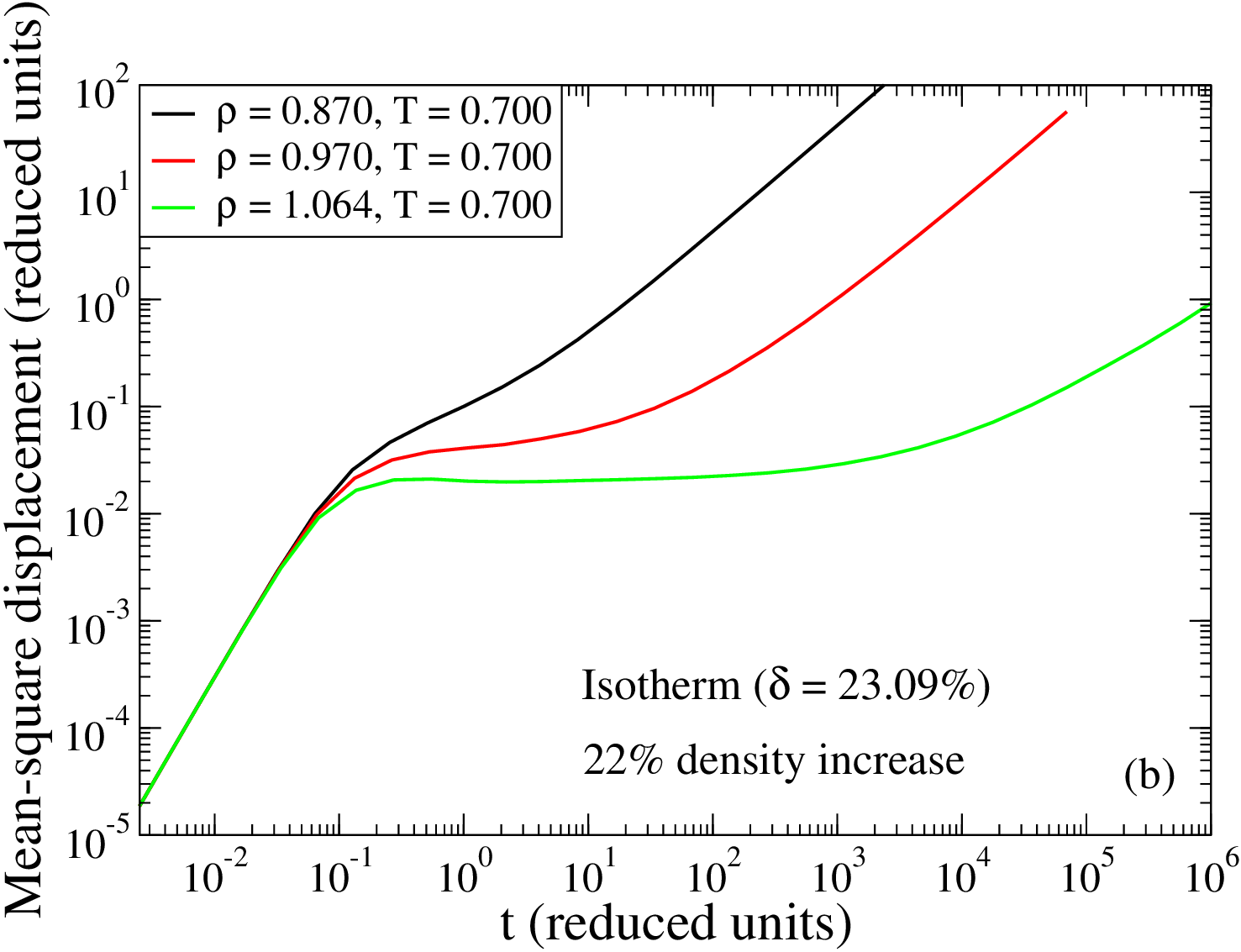}
  \caption{MSDs for a moderately polydisperse LJ liquid ($\delta$ =
    23.09\%) in three dimensions. (a) Along an isomorph. (b) Along an
    isotherm. The highest density state point 
    is not fully equilibrated.}
  \label{msd3d}
\end{figure}

Finally, the dynamics in terms of the MSD is shown in Fig. \ref{msd3d}. An excellent invariance is obtained along the isomorph, whereas the
isotherm shows approximately five orders-of-magnitude reduction in the diffusion coefficient.
\newline \newline 
\begin{figure}[H]
  \centering
  \includegraphics[width=70mm]{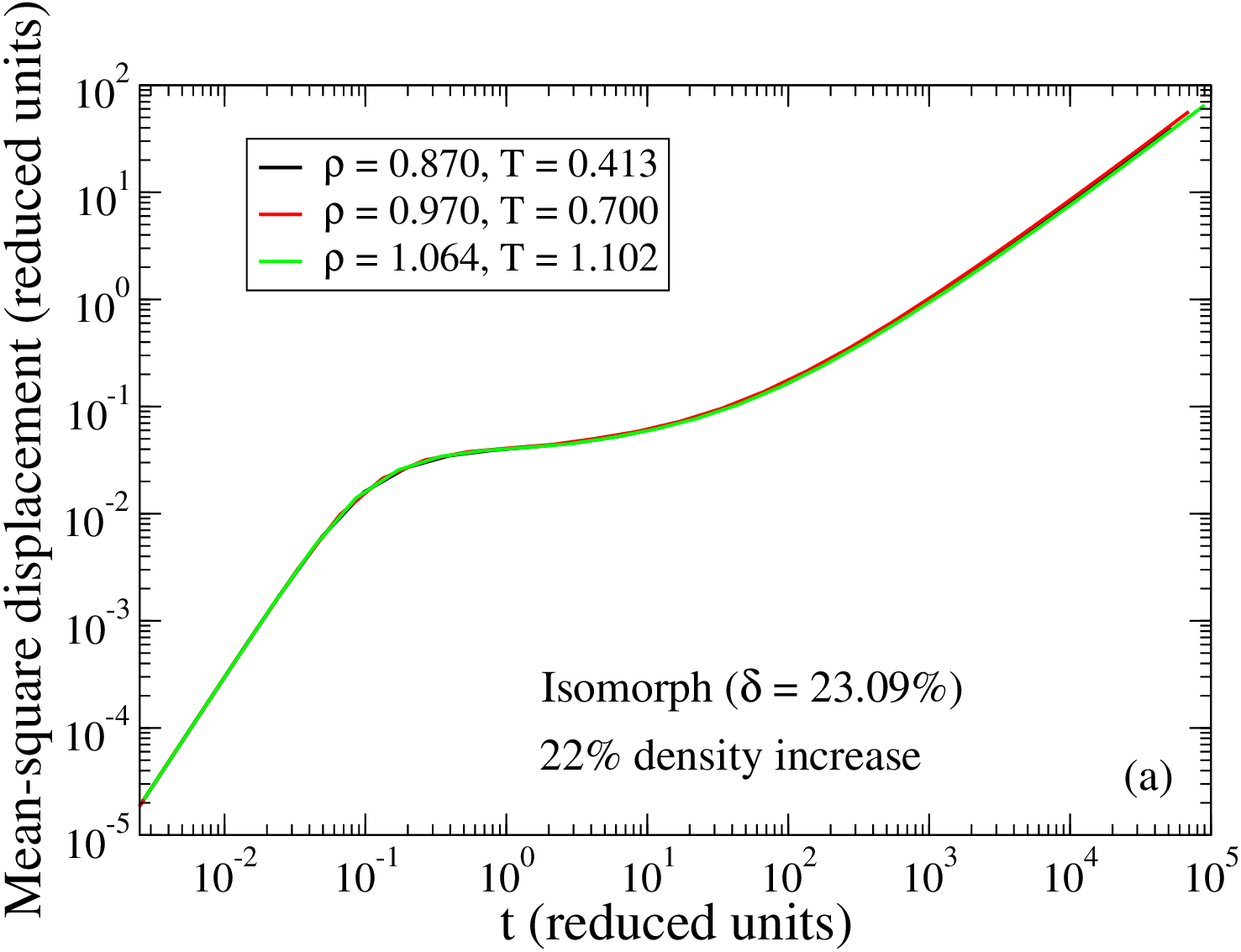}
\end{figure}

\section{Discussion}\label{dis}

We have demonstrated that both two- and three-dimensional size polydisperse LJ liquids belong to the class of RS liquids up to a 
significant level of polydispersity. This shows that the theory of isomorphs can be applied to multi-component systems 
which contributes to the simpler physical description of this important class of liquids. 
Nevertheless, our simulations also showed unusual behavior when
compared to the SCLJ liquid.
1) At higher densities, $R$ decreased with increasing polydispersity (pressure). An increase in pressure normally reduces interparticle
distances in which a pure IPL description is a better approximation, and
one would \textit{a priori} expect $R$ to increase. 2) For high polydispersities, the density-scaling exponent $\gamma$ is controlled only by the degree of
polydispersity exhibiting also significantly lower $\gamma$-values at
low-to-moderate pressures ($\gamma \approx 4.5$ versus $\gamma_{SCLJ}$
$\approx$ 5.5 in three dimensions). Below we try to understand these
observations in more detail presenting three-dimensional results only. 

\begin{figure}[h]
  \centering
  \includegraphics[width=70mm]{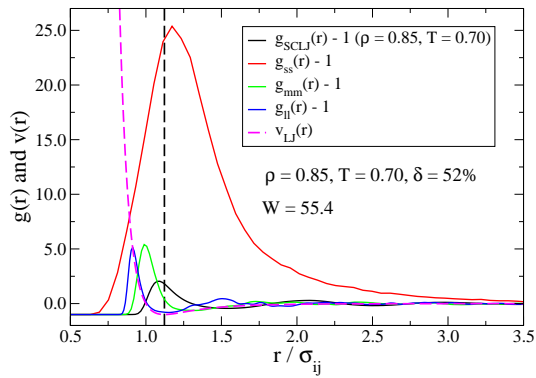}
   \caption{Like-like RDFs for the small ($s$), medium ($m$), and large ($l$) particles at $\rho$
     = 0.85, $T$ = 0.70, $\delta$ = 52\% (colored curves; $W$ =
     55.4) in three dimensions. $g_{ss}(r)$ uses particle sizes in the range $\sigma_{i}$ = $[0.1,0.2]$, 
    $g_{mm}(r)$ uses particle sizes in the range $\sigma_{i}$ = $[1.0,1.1]$, and $g_{ll}(r)$ uses particle sizes in the range $\sigma_{i}$ = $[1.8,1.9]$. 
    We note that this state point is not fully equilibrated. The black curve gives SCLJ liquid RDF 
    at $\rho$ = 0.85 and $T$ = 0.70 ($W$ = 0.31). The magenta dashed curve gives the LJ pair potential, and the black vertical dashed line delimits the minimum of the LJ potential.}
  \label{partialA}
\end{figure}

Figure \ref{partialA} shows like-like RDFs (colored
  curves) for the small, medium, and large particles at $\rho$ =
0.85, $T$ = 0.70, $\delta$ = 52\%, and the corresponding SCLJ liquid RDF (black curve). The RDFs for the medium and large particles move 
to shorter distances when compared to the SCLJ liquid (note the scaled distance axis). Conversely, the RDF for the small particles moves to larger distances and shows a very broad 
distribution. The higher value of the first peak is due to a clustering tendency of small particles in spaces surrounded by bigger particles 
(see Fig. \ref{2dplot})

The results signify a decoupling, with respect to pressure, amongst the small and large particles and could be the reason why $R$ decreases with 
increasing polydispersity (pressure) at higher densities. 
To verify this, we simulated at the same state point (not shown) a truncated 
uniform distribution excluding all the small(er) particles (i.e., particles with $\sigma_{i} < 1$). In this case: $R$ 
increased with polydispersity (pressure). 

Despite that the small particles are located around the minimum of the LJ potential, fluctuations 
in their positions are quite large. We speculate that this large-amplitude positional fluctuation
might diminish the impact of the particle-size dependent inhomogeneous stress distribution on $R$ (see Eq. (\ref{eq:potential})). 
Accordingly, it may also explain why the liquid remains RS up to such high polydispersities. 
Additional studies are, however, needed to clarify this issue. 
\newline \newline 

\begin{figure}[H]
  \centering
  \includegraphics[width=70mm]{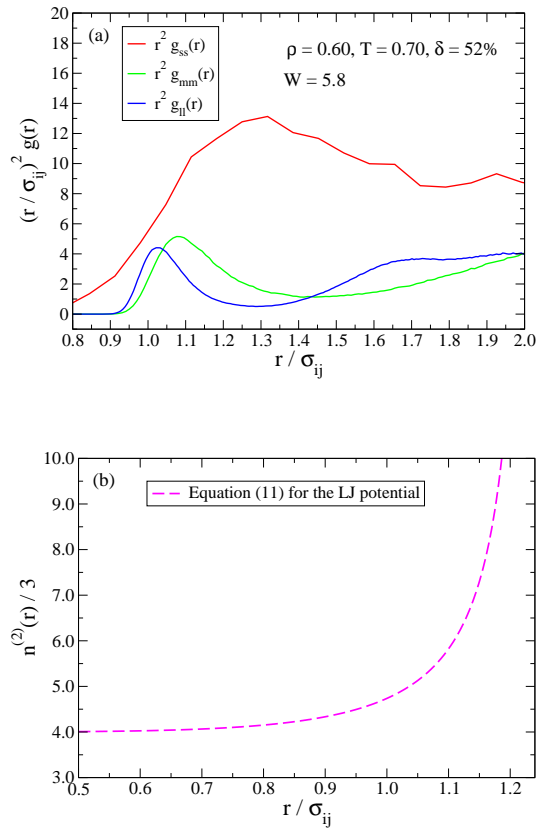}
  \caption{(a) $r^{2}g_{\alpha \alpha}(r)$ for the small ($s$), medium
    ($m$), and large ($l$) particles ($\alpha$ = $s$, $m$, $l$) at
    $\rho$ = 0.60, $T$ = 0.70, $\delta$ = 52\% ($W$ = 5.8) in three dimensions. $g_{ss}(r)$ uses particle sizes in the range $\sigma_{i}$ = $[0.1,0.2]$, $g_{mm}(r)$ uses particle sizes in the range $\sigma_{i}$ = $[1.0,1.1]$, 
    and $g_{ll}(r)$ uses particle sizes in the range $\sigma_{i}$ = $[1.8,1.9]$. (b) $n^{(2)}(r)$ of Eq. (\ref{n2a}) for the LJ potential.}
  \label{partialB}
\end{figure}

The investigation is now focused on 
 $\gamma$-values at low-to-moderate pressures for high polydispersities ($\gamma \approx 4.5$). In the light of the theoretical results on 
monatomic RS liquids (see Eq. (\ref{n2})), 
Fig. \ref{partialB}(a) shows $r^{2}g_{\alpha \alpha}(r)$ for the small ($s$), medium ($m$), and large ($l$) particles ($\alpha$ = $s$, $m$, $l$) 
at $\rho$ = 0.60, $T$ = 0.70, $\delta$ = 52\% ($W$ = 5.8). Figure \ref{partialB}(b) shows $n^{(2)}(r)$ for the LJ potential (Eq. (\ref{n2a})). Disregarding 
 now the fact that the liquid is polydisperse we apply $n^{(2)}(r)$ to make predictions 
for $\gamma_{\alpha}$ using the procedure mentioned in Sec. \ref{three} for RS monatomic liquids.  

In this way $\gamma_{m}$ $=$ 5.5 and $\gamma_{l}$ $=$ 4.9 are obtained for the medium 
and large particles, respectively, whereas $\gamma_{s} = -2.9$ gives a negative value due to 
the divergence in $n^{(2)}(r)$. Interestingly, $\gamma_{m}$ gives a value very close to that expected for a SCLJ liquid at low-to-moderate pressures ($\gamma_{SCLJ}$ $\approx$ 5.5).
Nevertheless, the results indicate that the large particles play an important role in determining the value of $\gamma$ for polydisperse LJ liquids with $\gamma_{l}$ $=$ 4.9. 
It is, however, unclear which role the small particles play, as $r^{2}g_{ss}(r)$ peaks at very large
distances with a resulting negative $\gamma_{s}$. 
We speculate that there is some cancellation between the medium particles of large $\gamma$ and the small particles of negative $\gamma$.
More in-depth investigations are, however, needed. 
\newline \newline 
\begin{figure}[h]
  \centering
  \includegraphics[width=70mm]{most_likely}
\end{figure}

\section{Conclusion and outlook}\label{con}

Polydisperse fluids show rich phenomenology with respect to their single-component fluid counterparts and manifest themselves in as diverse systems 
as bitumen in road construction\cite{hansen2013bit} to micelles in biological systems\cite{stapleton1988}. In spite of this fact, we have shown that 
even highly size polydisperse LJ liquids belong to the class of RS liquids. As a consequence, polydisperse LJ liquids inherit the 
associated simple properties of RS liquids\cite{prx,paper4} and are exemplified here by studying isomorphs of moderately polydisperse LJ liquids
in both two and three dimensions.

The SCLJ liquid is a prime example of a RS liquid, however, many more model and experimental liquids are identified to belong to this class 
of liquids. A natural question thus arises: How are the remaining
RS liquids affected by introducing size polydispersity?
As an example, identical chain length LJ-bead polymer models have very recently been identified as
RS\cite{veldhorst2014}. For polymers the relevant size polydispersity is given by a distribution of chain
lengths (mass distribution); can one expect such polymeric liquids to be RS? The answer is
most likely in the affirmative. It is known from experiments that scaling relations typical of RS liquids, such as density
scaling and excess entropy scaling, are also valid for polymers\cite{reviewRoland}. As another example, rigid-bond molecular liquids
with LJ interactions have also been identified as RS\cite{moleculesisomorphs}; it is expected
that these liquids remain RS when introducing size polydispersity. A more general hypothesis is that all RS liquids will remain so to a high degree of accuracy when 
introducing size polydispersity. Future research should, however, focus on clarifying to what extent this hypothesis is valid and on understanding the implications thereof. 

To conclude, the theory of isomorphs extends readily to multi-component systems, and when doing so it can be 
applied to improve even further the understanding of these intriguing
systems. Experimental liquids are often distributed in additional or other
variables such as mass or charge\cite{stapleton1988,ginoza1997,murarka2003}. More work in this direction is also needed to understand how these factors
influence the results presented here.

\acknowledgments

Valuable discussions with Lorentzo Costigliola, Jeppe C. Dyre, Claire Lemarchand, and Thomas B. Schr{\o}der are gratefully acknowledged.
This work was partially supported by Grant-in-Aid for Scientific Research (S) and Specially Promoted Research from the Japan Society
for the Promotion of Science (JSPS). T.S.I. acknowledges support from a JSPS Postdoctoral Fellowship.

\clearpage

\providecommand*\mcitethebibliography{\thebibliography}
\csname @ifundefined\endcsname{endmcitethebibliography}
  {\let\endmcitethebibliography\endthebibliography}{}


\end{document}